\documentclass[prl,superscriptaddress,aps,amsmath,floatfix]{revtex4}

\usepackage[normalem]{ulem}

\usepackage{graphics}
\usepackage{graphicx}
\usepackage{epstopdf}
\usepackage{dcolumn}
\usepackage{bm}
\usepackage{longtable}
\usepackage{epsfig}
\usepackage{times}
\usepackage{url}
\usepackage{color}

\begin{document}
\title{Synthetic gauge potentials for the dark state polaritons  in  atomic media}
\author{Yu-Hung   \surname{Kuan}}
\affiliation{Department of Physics, National Central University, Taoyuan City 32001, Taiwan}

\author{Siang-Wei   \surname{Shao}}
\affiliation{Department of Physics, National Central University, Taoyuan City 32001, Taiwan}

\author{I-Kang   \surname{Liu}}
\affiliation{Joint Quantum Centre Durham-Newcastle, School of Mathematics, Statistics and Physics, Newcastle University, Newcastle upon Tyne,  NE1 7RU, United Kingdom}

\author{Julius   \surname{Ruseckas}}
\affiliation{Baltic Institute of Advanced Technology, Pilies g. 16-8, LT-01403, Vilnius, Lithuania}
\affiliation{Institute of Theoretical Physics and Astronomy, Vilnius University, Saul\.{e}tekio 3, LT-10257 Vilnius, Lithuania}

\author{Gediminas    \surname{Juzeli\=unas}}
\affiliation{Institute of Theoretical Physics and Astronomy, Vilnius University, Saul\.{e}tekio 3, LT-10257 Vilnius, Lithuania}

\author{Yu-Ju   \surname{Lin}}
\affiliation{Institute of Atomic and Molecular Sciences, Academia Sinica, Taipei 10617, Taiwan}

\author{Wen-Te \surname{Liao}}
\email{wente.liao@g.ncu.edu.tw}
\affiliation{Department of Physics, National Central University, Taoyuan City 32001, Taiwan}
\affiliation{Physics Division, National Center for Theoretical Sciences, Hsinchu 30013, Taiwan}
\date{\today}
\begin{abstract}
The quest of utilizing neutral particles to simulate the behaviour of charged particles in a magnetic field  makes the generation of  artificial  magnetic field  of great interest.
The previous and the only proposal for the production of synthetic magnetic field for the dark state polaritons in electromagnetically induced transparency invokes the mechanical rotation of a sample.
Here, we put forward an optical scheme to generate effective gauge potentials for stationary-light polaritons.  To demonstrate the capabilities of our approach, we present recipes for having dark state polaritons in degenerate Landau levels and in driven quantum harmonic oscillator.
Our scheme paves a novel way towards the investigation of the bosonic analogue of the fractional quantum Hall effect by electromagnetically induced transparency.
\end{abstract}

\maketitle

\begin{figure}[b]
\includegraphics[width=0.5\textwidth]{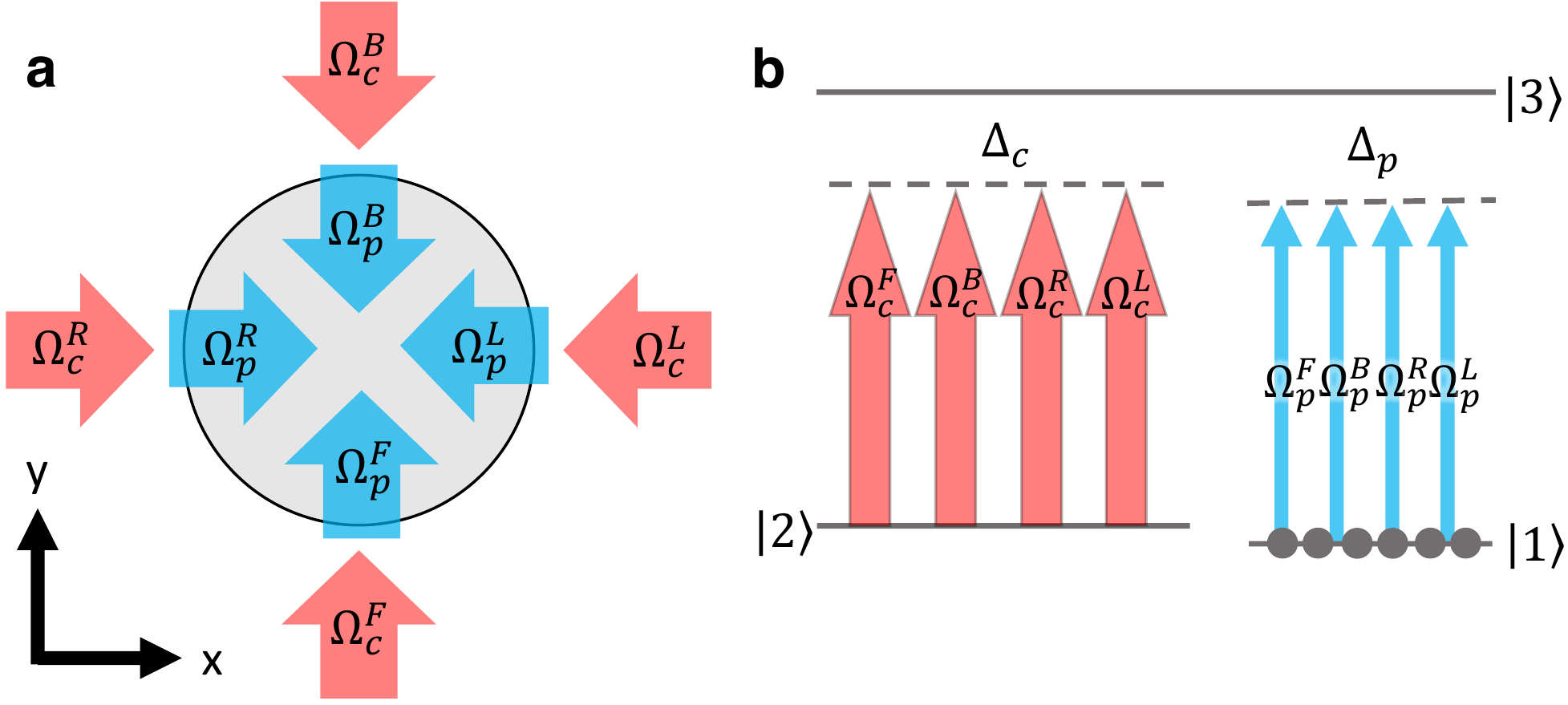}
\caption{\label{fig1}
{\bf Two-dimensional EIT system. } 
{\bf a}, The sketch for a two-dimensional EIT system.
{\bf b}, Three-level-$\Lambda$-type atomic system with two-photon detuning $\Delta_p-\Delta_c$.  Red (blue)-upward arrows represent the control (probe) fields, and their label denotes the Rabi frequency.
}
\end{figure}

Over the last decade  a considerable progress has been made in emulating the synthetic gauge fields for ultracold atoms \cite{Dalibard2011,Goldman2014,Lewenstein2012,Budich2016,Lin_2016,Cooper2019,Galitski2019}, photonic systems  \cite{Wang2015Optica,schine2016,Mukherjee2018NComm,Ozawa2019,clark2020,Marrucci2020}, and electric circuits \cite{Simon2020}. 
Among the photonic systems a special role is played by slow  \cite{Hau1999,fleischhauer2000,Juzeliunas2002,scully02PRL} and stationary \cite{bajcsy2003,Moiseev2006,Zimmer2008,lin2009} light forming in atomic media due to the electromagnetically induced transparency (EIT) \cite{Arimondo1996,Harris1997,Lukin2003,Fleischhauer2005,Bergman2017}. Such a light is composed of quasiparticles known as the dark state polaritons (DSPs) \cite{fleischhauer2000,Juzeliunas2002,Moiseev2006,Zimmer2008} made predominantly of atomic excitations, so the DSPs can interact strongly via the atom-atom interaction \cite{Gorshkov11PRL,Petrosyan11PRL,Lukin12Nature,Adams12ARCAP,gaerttner2014, Pohl17PRX,Firstenberg17RMP}. This can facilitate creating of strongly correlated quantum states, including the fractional Hall states. Yet the DSPs are electrically neutral quasiparticles and thus are not subjected  to the vector potential which provides the Lorentz force needed for the Hall effects. Up to now the only method considered for producing the synthetic gauge potential for the stationary light (stationary DSPs) involves rotation of the atomic medium \cite{Otterbach2010},  where a synthetic magnetic field is produced in the rotating frame. 
However, there are technical problems associated with synthetic fields in the rotating frame \cite{cornell2001, cooper2008}, and it is therefore desirable to engineer gauge potentials in the static laboratory frame.

In this article, we show a possible  optical method to engineer synthetic gauge potentials for stationary-light polaritons providing non-zero effective magnetic fields in the static laboratory frame. 
Therefore an EIT system  of  DSPs  can be a simulator for a charged  particle in  a magnetic field,  like ultracold atoms in the laser radiation \cite{Dalibard2011,Lewenstein2012,Goldman2014,Budich2016,Lin_2016,Cooper2019,Galitski2019,Juzeliunas2004,Juzeliunas2006,lin2009Sy}. We show a recipe to construct  environments for
Landau levels and a driven quantum harmonic oscillator
by engineering the synthetic vector and scalar potentials  for stationary DSPs.
The key ingredient of our idea is transferring the coupled Optical-Bloch equations (OBE) \cite{fleischhauer2000, bajcsy2003, lin2009} for a two-dimensional three-level-$\Lambda$-type EIT system (see Fig.~\ref{fig1}) to  
an electron-like Schr\"odinger equation for the dark-state  polarization $\rho_{21}$(see supplemental information)
\begin{equation}\label{eq1}
i\hbar\frac{\partial \rho_{21}}{\partial t} = \frac{\left( \frac{\hbar}{i}  \nabla + \vec{A} \right) ^2}{2m} \rho_{21}
+U\rho_{21} + i \left( \frac{\Gamma}{2\Delta_p}\right)\frac{\hbar^2}{2m} \nabla^2 \rho_{21},
\end{equation}
with $\hbar$ being the reduced Planck constant,  where the synthetic vector potential $\vec{A}$, and scalar potential energy $U$ read
\begin{eqnarray}
\vec{A} &=& m\vec{V}_g ,\label{eq2} \\
U &=& \hbar \left( \Delta_p-\Delta_c \right)  -  \frac{1}{2 m}\vert \vec{A} \vert^2. \label{eq3}
\end{eqnarray}
The dark-state  polarization  $\rho_{21}$ plays  a role of the wavefunction,  and 
the EIT group velocity  $\vec{V}_g = \left( V_R-V_L, V_F-V_B, 0 \right) = \frac{1 }{2\eta}\left( \vert \Omega_c^{R}\vert^2 - \vert \Omega_c^{L}\vert^2 ,\vert \Omega_c^{F}\vert^2 - \vert \Omega_c^{B}\vert^2, 0 \right)$ represents the vector potential $\vec{A}$ for a unit charge. 
Here $\Delta_p$ ($\Delta_c$) is the one-photon detuning of  the probe (control) fields, $\Omega_c^{F}$, $\Omega_c^{B}$, $\Omega_c^{R}$, and $\Omega_c^{L}$ are the Rabi frequencies of forward, backward, rightward and leftward propagating control fields, respectively,  with the same total intensities for the pairs of the couterpropagating beams
$\Omega_c^2=\vert \Omega_c^{R}\vert^2 + \vert \Omega_c^{L}\vert^2 = \vert \Omega_c^{F}\vert^2 + \vert \Omega_c^{B}\vert^2$,
%
%
%
$m=\frac{\hbar\eta^2}{2\Delta_p\Omega_c^2}$ is the effective mass when counter-propagating control fields are applied,  $\eta=\frac{\Gamma\xi_x}{2 L_x}=\frac{\Gamma\xi_y}{2 L_y}$  is the light-matter coupling constant,  $\Gamma$ is the spontaneous decay rate of the excited state $\vert 3\rangle$,  and 
$\xi_x$ ($\xi_y$) and $L_x$ ($L_y$) are  the optical depth and  the medium length in  the $x$ ($y$) direction, respectively. 
In equation \eqref{eq1} the kinetic energy term dominates over the last diffusion term  when $2\Delta_p\gg\Gamma$.

\section{Results}

\begin{figure*}[t]
\includegraphics[width=0.92\textwidth]{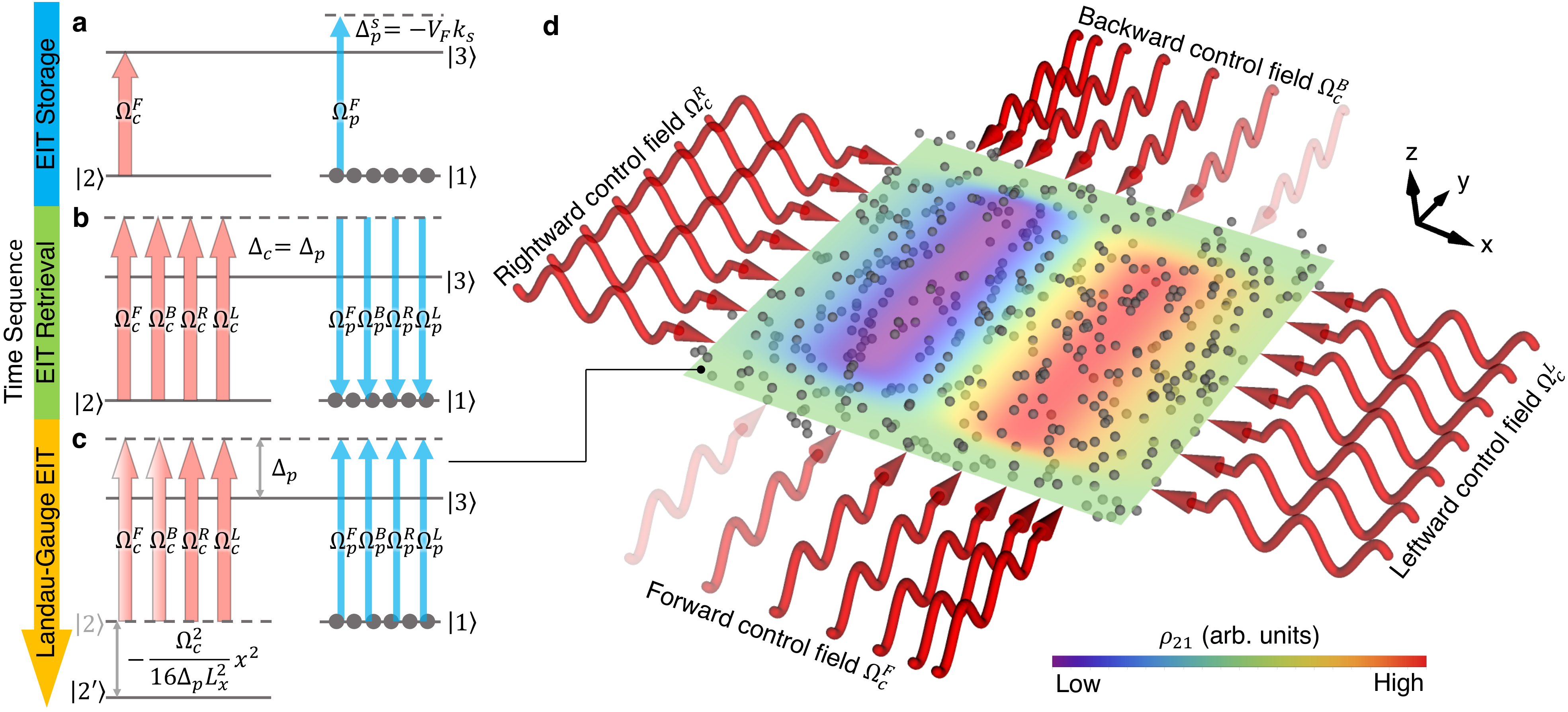}
\caption{\label{fig2}
{\bf Landau-gauge EIT system and time sequence. } 
 The time sequence is  
 {\bf a}, EIT light storage with probe one-photon detuning $\Delta_p^s$,  
 {\bf b}, retrieval, and  
 {\bf c}, switching on the  transversely gradient $\Omega_c^{F(B)}$ and the position-dependent two-photon detuning $\Delta_p-\Delta_c$.  
Red (blue)-upward arrows represent the control (probe) fields, and gray-vertical double arrows indicate detunings.
{\bf d}, The sketch for Landau-gauge EIT system. Gray dots represent atoms, and the red-sinusoidal arrows illustrate four control fields. The density and the opacity of arrows reflect the control field strength.
The coloured density plot depicts the spatial distribution of the dark-state  polarization $\rho_{21}$.
}
\includegraphics[width=0.92\textwidth]{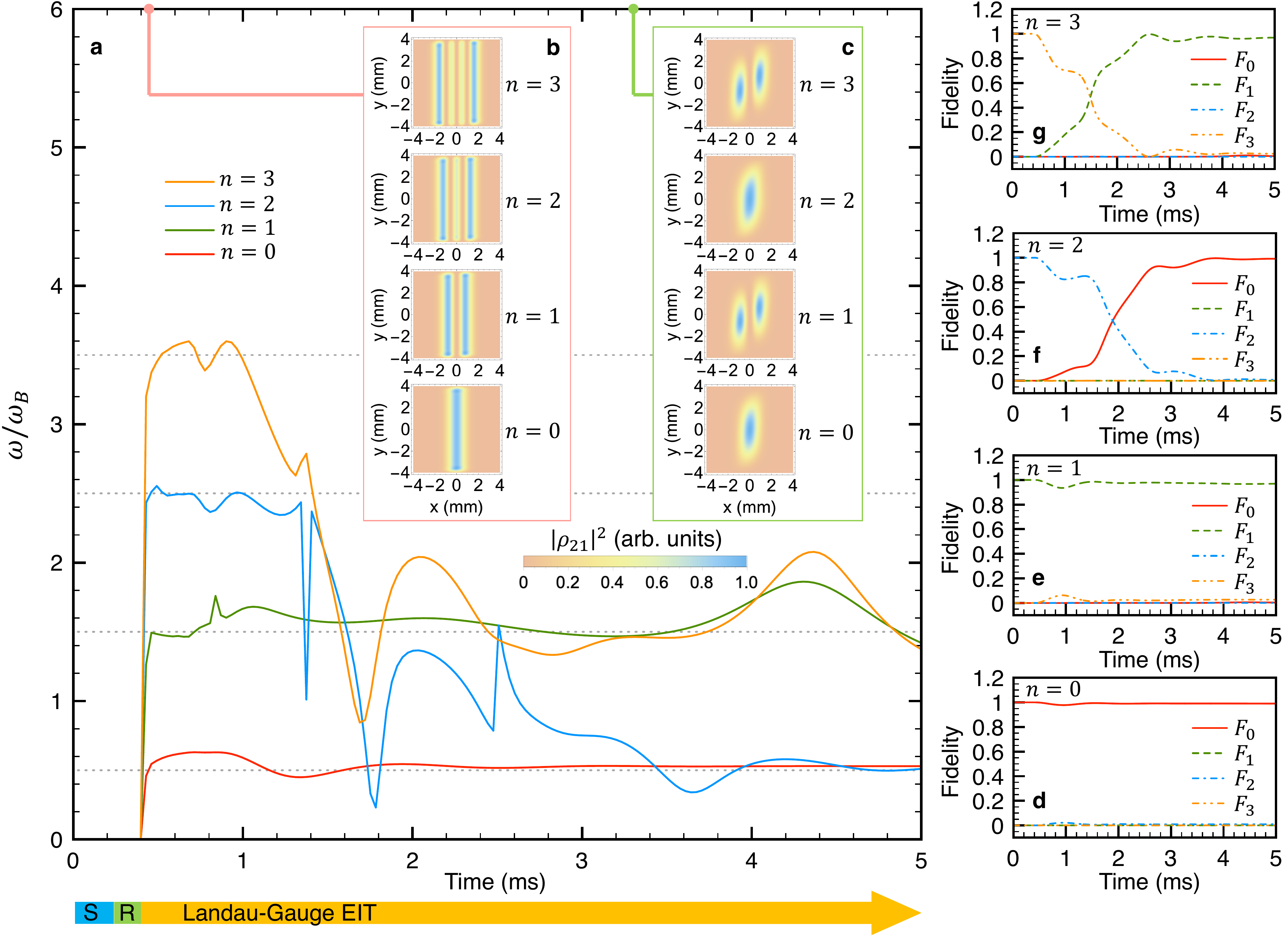}
\caption{\label{fig3}
{\bf The evolution of Landau levels. } 
 {\bf a}, The evolution of the angular frequency of the EIT ground state coherence $\rho_{21}$ initially prepared in the $n$th Landau level. 
Red-solid, green-solid, blue-sold, and orange-solid lines depict $n=0,1,2,3$, respectively. 
The bottom time arrow indicates different phases of the time sequence, where {\bf S} and {\bf R} indicate light storage and retrieval, respectively.
The spatial profile of $\vert \rho_{21} \vert^2$ in different Landau levels at {\bf b}, $t=0.44$ms when the Landau-gauge environment is switched on,  and at {\bf c}, $t=3.3$ms. The peak value of each $\vert \rho_{21} \vert^2$ is normalized to one.
 Fidelity for {\bf d}, $n=0$,
 {\bf e}, $n=1$,
 {\bf f}, $n=2$, and
 {\bf g}, $n=3$.
Red-solid, green-dashed, blue-dashed-dotted, and orange-dashed-dotted-dotted lines illustrate $F_0$, $F_1$, $F_2$, and $F_3$, respectively.
}
\end{figure*}
{\bf Landau levels.}
Given equations (\ref{eq1}-\ref{eq3}),  one can simulate a charged particle moving in a background uniform magnetic field, especially the so-called Landau levels,  in a two-dimensional EIT system. 
We manifest the generation of the synthetic $\vec{A}$ of Landau gauge for the EIT dark state by preparing $\rho_{21}$ in the $n$th Landau level $\vert n\rangle_L$ and then looking at its dynamics. 
Fig.~\ref{fig2}{\bf a}-{\bf c} depict the time sequence (see supplemental information for the complete time sequence)  used in our numerical simulations of  the full set of OBE.
First, in Fig. \ref{fig2}{\bf a} the initial state is prepared  by the typical EIT light storage technique  \cite{fleischhauer2000,Juzeliunas2002,Phillips2001,Liu2001}. 
A resonant uniform forward control field and a weak co-propagating  probe field with the transverse profile of the $n$th Landau level $\langle x\vert n\rangle_L $ and one-photon detuning $\Delta_p^{s} = -V_F k_s$  are injected into the medium along the $y$ axis.  By switching off the control field, the probe field is stored as the dark-state coherence in the medium
\begin{equation}
\rho_{21}\left( x,  y \right) = \rho_0 H_n\left(  \frac{x + k_s\mathit{l}_B^2}{\mathit{l}_B}\right) e^{ i k_s y  -\left(  \frac{x + k_s\mathit{l}_B^2}{\sqrt{2}\mathit{l}_B}\right)^2},
\end{equation}
where $H_n$ is the Hermite polynomial,   and $k_s$ is the  longitudinal wavenumber  of the slowly-varying amplitude.
%
%
%
%
Subsequently  four uniform control fields are turned on with  a certain  one photon detuning $\Delta_c$, and  the probe fields  is retrieved under the two-photon resonance condition $\Delta_p = \Delta_c$ as depicted by Fig. \ref{fig2}{\bf b}. 
This step endues $\rho_{21}$ with an  effective mass  by the homogeneous  detuning $\Delta_p$.
Finally,  we switch on the required control fields  with inhomogeneous detunings (gray-vertical double arrows) illustrated in Fig. \ref{fig2}{\bf c} to build the Landau-gauge environment: 
$\Delta_c = \Delta_p - \frac{ \Omega_c^2}{16 \Delta_p L_x^2} x^2$, $\Omega_c^R = \Omega_c^L = \frac{\Omega_c}{\sqrt{2}}$, $\Omega_c^F = \frac{\Omega_c}{\sqrt{2}}\sqrt{1+ \frac{x}{L_x}}$, and $\Omega_c^B = \frac{\Omega_c}{\sqrt{2}}\sqrt{1-\frac{x}{L_x}}$,  which  leads to $\vec{A} =  \frac{ \hbar \eta}{ 4 L_x \Delta_p} \left( 0, x, 0 \right)$  and $U=0$.
The inhomogeneous $\Delta_c$ can be implemented by  position-dependent Zeeman or Stark shifts to  displace the atomic level $\vert 2\rangle$ to $\vert 2'\rangle$ \cite{lin2009Sy, Otterbach2010}, as shown in Fig.~\ref{fig2}{\bf c}.
Fig.~\ref{fig2}{\bf d} illustrates our two-dimensional Landau-gauge EIT system.
The red-sinusoidal arrows denote the four control fields. The transversely gradient $\Omega_c^{F\left( B\right) }$ and the uniform $\Omega_c^{R\left( L\right) }$ are reflected by the density and opacity of arrows. Gray dots illustrate atoms, and the coloured density plot is the spatial profile of $\rho_{21}$.
With above choices,  the Landau levels are characterized by the magnetic length  
\begin{equation}\label{megneticlength}
\mathit{l}_B = \sqrt{\frac{8\Delta_p}{\xi_x\Gamma}}L_x =2\sqrt{\frac{\Delta_p L_x}{\eta}},
\end{equation}
and the cyclotron frequency 
$\omega_B = \frac{\Omega_c^2}{\xi_x\Gamma}$.
The  dynamics of $\rho_{21}$ on this stage is then governed by equation \eqref{eq1} giving three predictions:
(i) $\rho_{21} = \langle x\vert n\rangle_L$ evolves with the angular frequency of  $\omega_n=\left( n + 1/2\right) \omega_B$,  
(ii) the  decay from Landau level $\vert n+2\rangle_L$ to $\vert n\rangle_L$  takes place for $n\geq 2$ due to the last diffusion term (see supplemental information), and
(iii) the strip-like wavefunction centers at $x = -k_s\mathit{l}_B^2$ and extends in the $y$ direction.

Figure~\ref{fig3} demonstrates the first two predictions with $\Delta_p^{s} = 0$ and $k_s=0$. 
In  Fig. \ref{fig3}{\bf a} gray-dashed lines depict the theoretical  $\omega_n / \omega_B =\left( n + 1/2\right) $. The  angular frequency of $\rho_{21}$ is  numerically calculated by $\omega_n =i\left( \partial_t \rho_{21}\right)/\rho_{21} $ from our numerical solutions of OBE. 
The following EIT parameters are used:
$\Gamma=1$MHz,
$\Delta_p = 0.83 \Gamma$,
$\Omega_c = 1.5 \Gamma$,
$\xi_x = 900 $,
$\xi_y = 800 $,
$L_x = 9$mm, and
$L_y = 8$mm;  this results in 
$m=7\times 10^{-32}$kg,
$\omega_B=2.5$kHz, and
$\mathit{l}_B = 0.77$mm.
Red-solid, green-solid, blue-sold, and orange-solid lines depict  $\omega_n$ from our numerical solutions for $n=0,1,2,3$, respectively.
In the beginning at $t=0.44$ms when the required control fields and detunings for Landau gauge EIT are chronologically switched on, there are four branches of  the angular frequency, and each of  the numerically calculated $\omega_n$ matches the theoretical prediction very well. 
Later on  four  branches converge  in only two  bands which manifests the prediction (ii)  presented below Eq.\eqref{megneticlength}. 
For a better visualization of the Landau-level evolution, Fig~\ref{fig3}{\bf b} demonstrates the spatial distribution of $\vert\rho_{21}\vert^2$ for  a different input quantum number $n$ at $t=0.44$ms, and Fig~\ref{fig3}{\bf c} shows that at $t=3.3$ms. One can observe that each  Landau level $\vert n \geq 2\rangle_L$ gradually becomes $\vert n-2\rangle_L$, but $\vert n < 2\rangle_L$ sustains. The longitudinal asymmetry is caused by the finite size effect of $L_y$. 
Moreover, we calculate the fidelity 
$F_{n'}\left( t\right) =\vert _L\langle n' \vert \rho_{21}\left( t\right) \rangle  \vert^2/\langle \rho_{21}\left( t\right)  \vert \rho_{21}\left( t\right) \rangle$ to reveal the change of the projection of $\rho_{21}$ on the $n'$th Landau level $\vert n'\rangle_L$. The evolution of the fidelity for input $n=0,1,2$, and $3$ is illustrated in {\bf d}, {\bf e}, {\bf f}, and {\bf g}, respectively.
Remarkably, only $\Delta n =2$ spontaneous transitions $\vert 3 \rangle_L \rightarrow \vert 1 \rangle_L $ and $\vert 2 \rangle_L \rightarrow \vert 0 \rangle_L $ show up in Fig.~\ref{fig3}{\bf f}\&{\bf g}, and $\Delta n =1$ spontaneous decay is forbidden. This indicates the potential application of treating states $\vert 1 \rangle_L$ and $\vert 0 \rangle_L$ as a true and rather stable two-level system.

\begin{figure*}[b]
\includegraphics[width=0.59\textwidth]{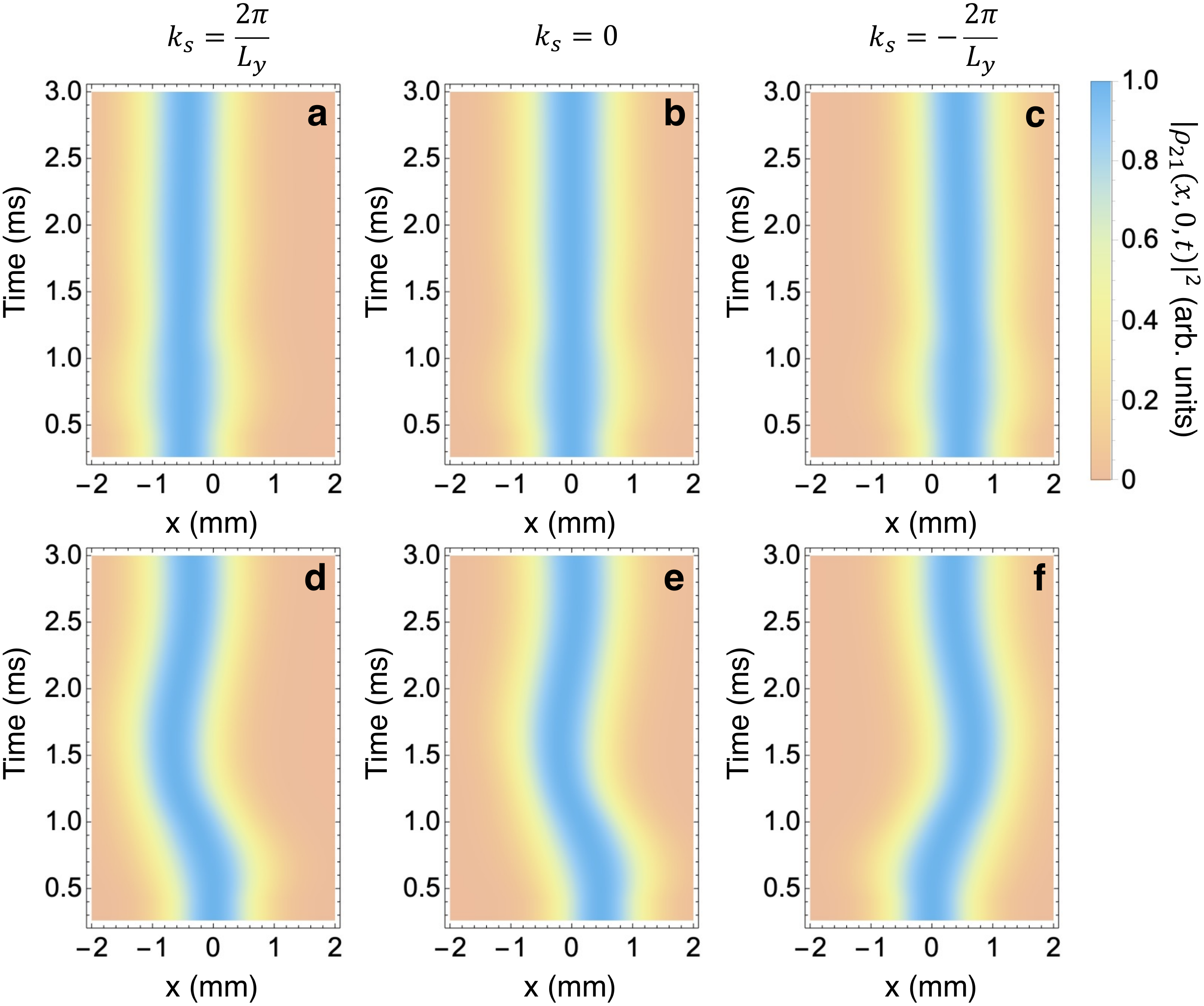}
\caption{\label{fig4}
{\bf $x_0$-dependent motion  of Landau levels. } 
The motion of $\vert \rho_{21}\left( x,0,t\right)  \vert^2$ for  
{\bf a}, $\left( k_s, x_0\right) =\left( \frac{2\pi}{L_y} , -k \mathit{l}_B^2\right) $, 
{\bf b}, $\left( k_s, x_0\right) =\left( 0 , 0\right)$,
{\bf c}, $\left( k_s, x_0\right) =\left( -\frac{2\pi}{L_y} , k \mathit{l}_B^2\right) $, 
{\bf d}, $\left( k_s, x_0\right) =\left( \frac{2\pi}{L_y} , 0\right)$,
{\bf e}, $\left( k_s, x_0\right) =\left( 0 , k \mathit{l}_B^2\right) $, and
{\bf f}, $\left( k_s, x_0\right) =\left( -\frac{2\pi}{L_y} , 0\right) $.
The peak value of $\vert \rho_{21}\left( t\right)  \vert^2$ is normalized to one.
}
\end{figure*}
\begin{figure*}[b]
\includegraphics[width=0.92\textwidth]{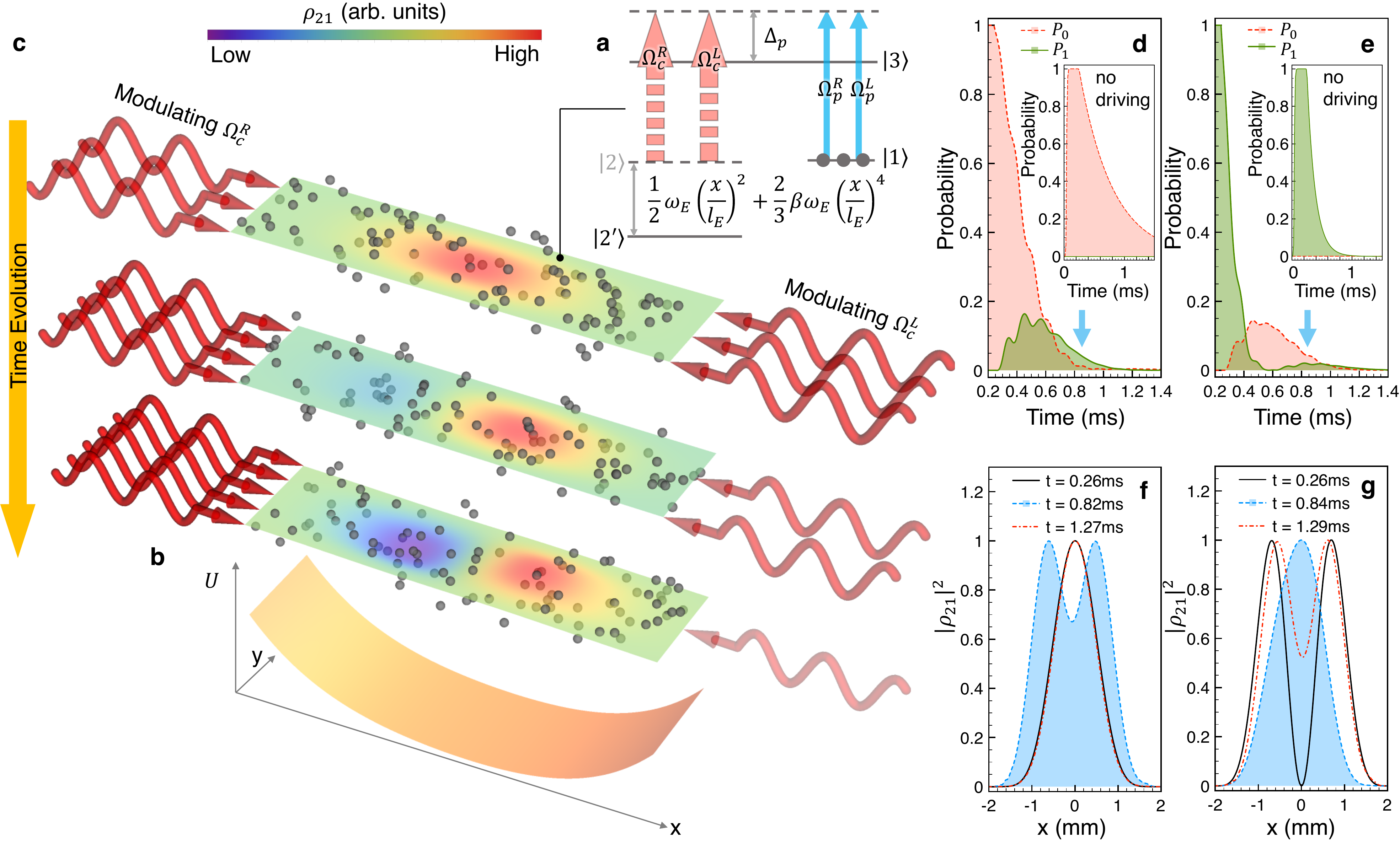}
\caption{\label{fig5}
{\bf EIT simulator for a driven quantum harmonic oscillator. }  
{\bf a}, Three-level EIT system. Blue-upward arrows depict the probe fields. The red-thick-dashed-upward arrows denote the periodically modulating $\Omega_c^R$ and $\Omega_c^L$. The shifted level $\vert 2'\rangle$ indicates the position-dependent two-photon detuning $\Delta_p-\Delta_c$.  
{\bf b}, The $x$-dependent synthetic scalar potential energy $U$.
{\bf c}, The sketch of the EIT simulator for a driven QHO. Gray dots represent atoms, and the red-sinusoidal arrows illustrate two counter-propagating control fields. The density and the opacity of arrows reflect the control field strength.
The coloured density plot depicts the profile of the coherence $\rho_{21}$. 
The evolution of the state probability $P_{n'} $ of the driven two-level system for initial states  {\bf d},  $\vert 0 \rangle$ and  {\bf e},  $\vert 1 \rangle$. Each inset is the result for a pair of static $\Omega_c^R=\Omega_c^L$.
The snapshots of $\vert\rho_{21}\left( x, t\right) \vert^2$ for the driven transition {\bf f}, $\vert 0 \rangle \rightarrow \vert 1 \rangle$ and  {\bf g},  $\vert 1 \rangle \rightarrow \vert 0 \rangle$.
}
\end{figure*}
The prediction (iii) is one of the remarkable properties of  the Landau level $\langle x-x_0\vert n\rangle_L e^{ik_s y}$. One can test the Landau-gauge EIT via observing the $x_0$-dependent motion of $\rho_{21}$ for a given  $k_s$. 
To prepare the initial state for  the above purpose (see supplemental information for the complete time sequence),  as depicted by Fig.~\ref{fig2}{\bf a}, one can store a probe field with the transverse profile $\langle x-x_0\vert n\rangle_L $ centering on $x=x_0$ and non-zero $\Delta_p^{s} = -V_F k_s$.
Figures~\ref{fig4}{\bf a}-{\bf c} illustrate the stationary motion of $\vert \rho_{21}\left( x, y=0, t\right)  \vert^2$ for $k_s =\frac{2\pi}{L_y}, 0$ and $-\frac{2\pi}{L_y}$, respectively. Three wavenumbers are accordingly produced by $\Delta_p^{s} = -0.018\Gamma$, $\Delta_p^{s} = 0$, and $\Delta_p^{s} = 0.018\Gamma$.
Our numerical solutions of OBE show the motional stability at  their exact predicted  peak position $x_0=-k_s \mathit{l}_B^2$. 
The peak value of $\vert \rho_{21}\left(  t\right)  \vert^2$ is normalized to  the unity for the sake of better visualization.
On the other hand, having  $x_0\neq -k_s \mathit{l}_B^2$ causes  a snake-like  motion,  as demonstrated in  Figs.~\ref{fig4}{\bf d}-{\bf f} for $k_s =\frac{2\pi}{L_y}, 0$ and $-\frac{2\pi}{L_y}$, respectively.  This reflects the coherent state oscillation when the center of the Landau-gauge harmonic trap is shifted to $x = -k_s \mathit{l}_B^2$. Each non-stationary coherent state gradually decays to the ground state also due to the diffusion term in equation \eqref{eq1}.
The $x_0$-dependent motion of  the Landau level corresponds to the shifted harmonic potential and fulfils the prediction (iii) by equation \eqref{eq1}. The dynamics of Landau levels reveals the effects of synthetic vector potential on  the DSPs.

As a comparison with the mechanically rotating method \cite{Otterbach2010}, we estimate the magnetic length of two schemes. 
In  the case of Eq.~\eqref{megneticlength}, when the transverse slope of the control fields  approaches the diffraction limit, i.e., $L_x\approx\lambda$, and the kinetic energy term dominates over the diffusion one with the conservative choice of $\Delta_p = 10\Gamma$ in Eq.~\eqref{eq1},  we obtain  the minimum $\mathit{l}_{B} = \frac{1}{\sqrt{\eta}}\sqrt{40 \lambda\Gamma}$. 
In the rotating scheme  one  obtains the magnetic length $\mathit{l}_{B}^{rot} = \sqrt{\frac{\lambda}{4\pi}\frac{V_g}{\nu}} = \frac{1}{\sqrt{\eta}} \sqrt{\frac{\lambda\Omega_c^2}{8\pi\nu}}$, where $V_g$ is the EIT slow light group velocity and $\nu$ the sample rotating angular frequency. 
The following typical values $\nu=1$kHz $=10^{-3}\Gamma$ and $\Omega_c=1\Gamma$ also result in   
$\mathit{l}_{B}^{rot}  = \frac{1}{\sqrt{\eta}}\sqrt{40 \lambda\Gamma}$.
Accordingly, our optical scheme can generate a synthetic magnetic field  similar  to that obtained by the rotation of a sample. 
The degeneracy of the lowest Landau level (LLL)  is 
$ D_{lll} = \frac{L_x L_y}{2\pi \mathit{l}_B^2}$, and the filling factor is 
$\nu_\mathrm{filling} = 2\pi n_\mathrm{D}\mathit{l}_B^2$,
where $n_\mathrm{D}$ is  the concentration of two-dimensional  DSPs \cite{Otterbach2010}.
Given  above estimation $\mathit{l}_B \geq \frac{1}{\sqrt{\eta}}\sqrt{40 \lambda\Gamma}$, we obtain two conditions
$D_{lll} \leq \frac{L_x L_y \eta}{80\pi \lambda\Gamma}$ and
$\nu_\mathrm{filling} \geq \frac{80\pi n_\mathrm{D}\lambda\Gamma}{\eta}$.
With the set of parameters used in Fig.~\ref{fig3} and  $\lambda = 500$nm, one gets the maximum $D_{lll}= 2.9\times 10^4$.
A careful arrangement of the input probe photon distribution  leads to $n_\mathrm{D}=\frac{D_{lll}}{M L_x L_y}=\frac{ \eta}{80 M \pi \lambda\Gamma}$, where  $M$ can be some integer \cite{sorensen2005, Hafezi2007, Sterdyniak2012}.  For $M > 1$ this results in  a  filling  $\nu_\mathrm{filling} = 1/M$  smaller than one.  The vector potential $\vec{A}$ can  be also engineered  in   the symmetric gauge by introducing  the gradients of the strength   to four control fields.

{\bf Driven quantum harmonic oscillator.}
We now turn to simulate a driven quantum harmonic oscillator (QHO) with EIT by 
using synthetic stationary $U$ and time-varying $\vec{A}$.
When mimicking the driven QHO  system, a  time sequence similar to  that shown in Figs.~\ref{fig2}{\bf a}-{\bf c} can be utilized (see supplemental information for the complete time sequence).
The initial state  is prepared by storing a probe field with the profile of $\langle x\vert n\rangle_H$ where $\vert n\rangle_H$ is the $n$th   eigenstate of QHO.
This generates
\begin{equation}\label{eq9}
\rho_{21}\left( x \right)  = \rho_0 H_n\left(  \frac{x}{\mathit{l}_E}\right) e^{-\left(  \frac{x}{\sqrt{2}\mathit{l}_E}\right)^2}
\end{equation} 
in the medium.
Afterwards, we turn on two counter-propagating control fields $\Omega_c^R = \Omega_c^L = \frac{\Omega_c}{\sqrt{2}}$ 
to retrieve a stationary pulse \cite{bajcsy2003, lin2009, everett2017, park2018} under the two-photon resonance condition $\Delta_c = \Delta_p $.
Subsequently, we switch on the one-photon detuning $\Delta_c = \Delta_p  - \frac{1}{2}\omega_E \left(\frac{x}{\mathit{l}_E}\right)^2-\frac{2}{3}\beta\omega_E \left(\frac{x}{\mathit{l}_E}\right)^4$
depicted in the three-level scheme by Fig. \ref{fig5}{\bf a}. 
On this stage $\rho_{21}$ evolves as the $n$th QHO eigenstate in a harmonic trap under the quartic perturbation, namely, a synthetic $U=\frac{1}{2}m\omega_E^2 x^2+\beta\frac{2 m^2 \omega_E^3}{3\hbar} x^4$ illustrated in Fig. \ref{fig5}{\bf b}. 
The characteristic QHO length is given by $\mathit{l}_E = \frac{\Omega_c}{\xi_x\Gamma}\sqrt{\frac{8\Delta_p}{\omega_E}}L_x $.
In view of equation \eqref{eq1}, we can introduce
$\Omega_c^R=\frac{\Omega_c}{\sqrt{2}}\sqrt{1+\alpha\sin\left( \omega_d t\right) }  $, and $\Omega_c^L=\frac{\Omega_c}{\sqrt{2}}\sqrt{1-\alpha\sin\left( \omega_d t\right) }$
to construct
$\vec{A}=\left( \alpha\frac{\hbar\eta}{4\Delta_p} \sin\left( \omega_d t\right), 0, 0\right)  $.
The periodic modulation of the control fields shakes the dark-state polarization in a  trapping potential $U$ as demonstrated in Fig. \ref{fig5}{\bf c}. 
Note that we are aiming  at simulating the Hamiltonian neglecting the $\vert\vec{A}\vert^2$ term in Eqs.~(\ref{eq1}\&\ref{eq3}) as typically adopted in quantum optics, and so the two-photon detuning is not dynamically modulated.
We derive the effective driving Rabi frequency $\Omega_A=\frac{1}{\hbar} {_H\langle n}+1\vert \frac{\hbar e}{i m}\vec{A}\cdot\nabla \vert n\rangle_H = \frac{\alpha\Omega_c}{4}\sqrt{\frac{\left( n+1\right) \omega_E}{\Delta_p}}$ and the perturbed eigen angular frequency $\omega_n = \left( n+\frac{1}{2}\right) \omega_E + \beta \left( n^2 + n +\frac{1}{2}\right)\omega_E $ for $\beta\ll 1$.
The quartic potential breaks the equal energy spacing between neighbouring QHO eigenstates and renders driving only a chosen dipole transition possible by matching $\omega_d = \omega_{n+1}-\omega_n$.
Fig.~\ref{fig5}{\bf d}-{\bf g} depict our numerical results with
the following EIT parameters:
$\alpha=0.24$,
$\beta=0.15$,
$\omega_E=20$rad$\cdot$kHz,
$\omega_d=26$rad$\cdot$kHz,
$\Gamma=1$MHz,
$\Delta_p = 4.6 \Gamma$,
$\Omega_c = 1.5 \Gamma$,
$\xi_x = 800 $, 
and
$L_x = 8$mm, which
lead to
$m = 1.27\times 10^{-32}$kg, 
$\mathit{l}_E = 0.644$mm, and
$\Omega_A = 6$rad$\cdot$kHz for the  QHO $\vert 0\rangle_H\rightarrow\vert 1\rangle_H$ transition.
We calculate the state probability
$P_{n'}\left( t\right) =\vert {_H\langle} n' \vert \rho_{21}\left( t\right) \rangle  \vert^2/\langle \rho_{21}\left( t_0\right)  \vert \rho_{21}\left( t_0\right) \rangle$, where the dynamical  modulation of $\Omega_c^R$ is switched on at $t=t_0 = 0.26$ms and show the case for initial state $\rho_{21}\left( x, t_0\right) = \langle x\vert 0\rangle_H$ in Fig.~\ref{fig5}{\bf d}. 
It is clear to see that the occurrence of  $\vert 0\rangle_H\rightarrow\vert 1\rangle_H$ transition as predicted by equation \eqref{eq1}. $P_0$ takes about 0.65ms  to drop from 1 to 0, and $P_1$ simultaneously grows. 
As a comparison, the inset shows the case without dynamical modulation. The free decay of $P_0$ and non-growth $P_1$ reflect the dissipation of the EIT system in state $\vert 0\rangle_H$, and 
$P_0$  spends about 2ms to descend from 1 to 0.
The coherently driven excitation rate is significantly greater than the free decay rate of state $\vert 0\rangle_H$. 
Moreover, we show the stimulated deexcitation of $\vert 1 \rangle_H \rightarrow \vert 0 \rangle_H $ in Fig.~\ref{fig5}{\bf e}, where initial state $\vert 1\rangle_H$ is prepared. 
Compared with the inset  using a pair of static $\Omega_c^R$ and $\Omega_c^L$,
not only the speed-up deexcitation due to the dynamical modulation happens, but also the revival of the first excited state occurs at $t=0.9$ms. The latter is the signature of the damped Rabi oscillation in strong coupling regime, i.e., $\Omega_A \geq$ the decay rates.
In order to further visualize the strong coupling effect, we illustrate $\vert \rho_{21}\vert^2$ at $t=0.26$ms, 
$t=0.82$ms 
, and 
$t=1.27$ms 
for the initial state $\vert 0 \rangle_H $ in Fig.~\ref{fig5}{\bf f}. Also, we show $\vert \rho_{21}\vert^2$ at $t=0.26$ms, $t=0.84$ms, and $t=1.29$ms for the initial state $\vert 1 \rangle_H $ in Fig.~\ref{fig5}{\bf g}. 
The clear alternation between $\vert\langle x\vert 0 \rangle_H \vert^2$ and $\vert\langle x\vert 1 \rangle_H \vert^2$ reveals the strong coherent coupling between two states as predicted by equation \eqref{eq1}. The oscillation period is very close to the theoretical prediction, e.g., the second instants indicated by blue downward arrows in Fig.~\ref{fig5}{\bf d}\&{\bf e} are near $0.26$ms + $\pi/\Omega_A\approx 0.78$ms.

\section{Discussion}
We have put forward a possible optical method to generate synthetic gauge fields for neutral EIT DSPs. 
Our optical scheme can not only produce comparable artificial magnetic field to that by mechanical rotation \cite{Otterbach2010} but also  provides a versatile platform to simulate different Hamiltonians, e. g., the demonstrated QHO  for DSPs.
Moreover, highly degenerate Landau levels are expected to be prepared by carefully adjusting the distribution of the number of interacting DSPs among LLL strips.
The  above DSP's dynamics can be observed by the direct imaging technique \cite{campbell2017} or by the retrieving of the  probe fields at different directions like tomography. An optical depth over 1000 has been experimentally achieved \cite{blatt2014, hsiao2018}. 
Our scheme is also novel  for the investigation of controllable non-Hermitian quantum systems.

\section{Methods}
The probe light propagating in four directions are simulated by matrix method  which stems from the Crank-Nicolson method. The behavior of atoms is simulated by the fourth-order Runge-Kutta method. The size of each temporal grid is about $ 300ns $, and the spacial grid is about $ 0.1mm $.

\section{Acknowledgements}
This work is supported by the Ministry of Science and Technology, Taiwan (Grant No. MOST 107-2112-M-008-007-MY3,  MOST 109-2639-M-007-002-ASP \& MOST 106- 2112-M-018-005-MY3). 
I-K.L. was supported by the  Quantera ERA-NET cofund project NAQUAS through the Engineering and Physical Science Research Council, Grant No. EP/R043434/1.
G.J.  and J.R.  were also supported by the National Center for Theoretical Sciences, Taiwan. 

\section{Author contributions}
Y.-H. K. programmed the computational code. 
Y.-H. K. and S.-W. S. performed the numerical calculations.
W.-T. L. and I.-K. L.  derived the physics model.
W.-T. L., Y.-J. L.,  and G. J. conceived the idea
W.-T.~L.,  conducted the project. 
All the authors discussed the results and wrote the manuscript.

%
\bibliographystyle{apsrev}
\bibliography{LGEIT}

\section{Synthetic gauge potentials for the dark state polaritons  in  atomic media: supplemental information}

\maketitle
The detail of our derivations and the time sequences used in our numerical simulation are demonstrated.
We begin with Maxwell-Schr\"odinger equation in perturbation region, namely, $\vert\Omega_c \vert\gg \vert\Omega_p\vert$ \cite{Scully2006, Lin2009} and follow the derivation from \cite{Fleischhauer2000}.
\begin{equation}\label{eq1}
\frac{\partial \rho_{21}}{\partial t} = 
\frac{i}{2} \Omega_c^{F*} \rho_{31}^F + 
\frac{i}{2} \Omega_c^{B*} \rho_{31}^B +
\frac{i}{2} \Omega_c^{R*} \rho_{31}^R +
\frac{i}{2} \Omega_c^{L*} \rho_{31}^L +
i \left( \Delta_c - \Delta_p\right)   \rho_{21},
\end{equation}
and
\begin{eqnarray}
\frac{\partial \rho_{31}^F}{\partial t} &=& \frac{i}{2} \Omega_p^F + \frac{i}{2} \Omega_c^F \rho_{21} -\left(  \frac{\Gamma}{2} +i\Delta_p \right) \rho_{31}^F,\label{eq2}\\
\frac{\partial \rho_{31}^B}{\partial t} &=& \frac{i}{2} \Omega_p^B + \frac{i}{2} \Omega_c^B \rho_{21} -\left(  \frac{\Gamma}{2} +i\Delta_p \right)  \rho_{31}^B,\label{eq3}\\
\frac{\partial \rho_{31}^R}{\partial t} &=& \frac{i}{2} \Omega_p^R + \frac{i}{2} \Omega_c^R \rho_{21} -\left(  \frac{\Gamma}{2} +i\Delta_p \right)  \rho_{31}^R,\label{eq4}\\
\frac{\partial \rho_{31}^L}{\partial t} &=& \frac{i}{2} \Omega_p^L + \frac{i}{2} \Omega_c^L \rho_{21} -\left(  \frac{\Gamma}{2} +i\Delta_p \right)  \rho_{31}^L,\label{eq5}
\end{eqnarray}
and wave equations
\begin{eqnarray}
\frac{1}{c} \frac{\partial \Omega_p^F}{\partial t} + \frac{\partial \Omega_p^F}{\partial y} &=& i \eta \rho_{31}^F+\frac{i}{2 k}\nabla^2\Omega_p^F ,\label{eq6}\\
\frac{1}{c} \frac{\partial \Omega_p^B}{\partial t} - \frac{\partial \Omega_p^B}{\partial y} &=& i \eta \rho_{31}^B+\frac{i}{2 k}\nabla^2\Omega_p^B ,\label{eq7}\\
\frac{1}{c} \frac{\partial \Omega_p^R}{\partial t} + \frac{\partial \Omega_p^R}{\partial x} &=& i \eta \rho_{31}^R+\frac{i}{2 k}\nabla^2\Omega_p^R ,\label{eq8}\\ 
\frac{1}{c} \frac{\partial \Omega_p^L}{\partial t} - \frac{\partial \Omega_p^L}{\partial x} &=& i \eta \rho_{31}^L+\frac{i}{2 k}\nabla^2\Omega_p^L \label{eq9}.
\end{eqnarray}
Here $\Omega^{F \left( B, R, L\right) }_c$ is the Rabi frequency of forward (backward, rightward, leftward) control field, and $\Omega^{F \left( B, R, L\right) }_p$ is the Rabi frequency of forward (backward, rightward, leftward) probe field. $\rho_{21}$ is the ground state coherence between state $\vert 1\rangle$ and $\vert 2\rangle$, and $\rho^{F \left( B, R, L\right) }_{31}$ is the coherence between state $\vert 1\rangle$ and $\vert 3\rangle$ for forward ( backward, rightward, leftward ) EIT configuration. $\Delta_c$ and $\Delta_p$ are the detuning of control field and that of probe laser, respectively.

The basic strategy of derivation is to express all terms of probe field by $\rho_{21}$ under adiabatic condition.
First, we neglect $\frac{\partial \rho_{21}}{\partial t}$ and $\frac{\partial \rho^{F \left( B, R, L\right) }_{31}}{\partial t}$  in Eq.~(\ref{eq1}-\ref{eq5}) and then get
\begin{eqnarray}
\rho_{21} &=& \frac{-\left( \Omega_c^{F*}\Omega_p^{F}+\Omega_c^{B*}\Omega_p^{B}+\Omega_c^{R*}\Omega_p^{R}+\Omega_c^{L*}\Omega_p^{L}\right) }{2\left( \Delta_c - \Delta_p\right)\left( 2\Delta_p - i\Gamma \right) + \vert \Omega_c^{F}\vert^2+\vert \Omega_c^{B}\vert^2+\vert \Omega_c^{R}\vert^2+\vert \Omega_c^{L}\vert^2},\label{eq10}\\
\Omega_p^{F} &=& - \Omega_c^{F} \rho_{21} + \left( 2\Delta_p - i \Gamma \right)\rho_{31}^F  
\sim -\left[ 1+ \left( \frac{ 2\Delta_p - i \Gamma }{i\eta}\right) \frac{\partial }{\partial y} \right]  \Omega_c^{F} \rho_{21},\label{eq11}\\ 
\Omega_p^{B} &=& - \Omega_c^{B} \rho_{21} + \left( 2\Delta_p - i \Gamma \right)\rho_{31}^B 
\sim -\left[ 1- \left( \frac{ 2\Delta_p - i \Gamma }{i\eta}\right) \frac{\partial }{\partial y} \right]  \Omega_c^{B} \rho_{21},\label{eq12}\\ 
\Omega_p^{R} &=& - \Omega_c^{R} \rho_{21} + \left( 2\Delta_p - i \Gamma \right)\rho_{31}^R
\sim -\left[ 1+ \left( \frac{ 2\Delta_p - i \Gamma }{i\eta}\right) \frac{\partial }{\partial x} \right]  \Omega_c^{R} \rho_{21},\label{eq13}\\ 
\Omega_p^{L} &=& - \Omega_c^{L} \rho_{21} + \left( 2\Delta_p - i \Gamma \right)\rho_{31}^L
\sim -\left[ 1- \left( \frac{ 2\Delta_p - i \Gamma }{i\eta}\right) \frac{\partial }{\partial x} \right]  \Omega_c^{L} \rho_{21}.\label{eq14}
\end{eqnarray}
In order to get the last term in Eq.(\ref{eq11}-\ref{eq14}), we invoke $\vert \Omega^{F \left( B, R, L\right) }_c \rho_{21}\vert \gg \vert \left( 2\Delta_p - i \Gamma \right)\rho_{31}^{F \left( B, R, L\right) }  \vert$ and neglecting time derivative and $\nabla^2$ in Eq.~(\ref{eq6}-\ref{eq9}).
Equation~(\ref{eq1}) also leads to
\begin{equation}\label{eq15}
\Omega_c^{F*} \rho_{31}^F + 
\Omega_c^{B*} \rho_{31}^B +
\Omega_c^{R*} \rho_{31}^R +
\Omega_c^{L*} \rho_{31}^L 
=\frac{2}{i}\left[ \frac{\partial }{\partial t}- i \left(  \Delta_c - \Delta_p  \right)  \right]  \rho_{21}
\end{equation}
By doing 
$\Omega_c^{F*}\times$Eq.~(\ref{eq6})$+\Omega_c^{B*}\times$Eq.~(\ref{eq7})$+\Omega_c^{R*}\times$Eq.~(\ref{eq8})$+\Omega_c^{L*}\times$Eq.~(\ref{eq9}), one gets
\begin{eqnarray}\label{eq16}
&&\frac{1}{c} \frac{\partial \left( \Omega_c^{F*}\Omega_p^{F}+\Omega_c^{B*}\Omega_p^{B}+\Omega_c^{R*}\Omega_p^{R}+\Omega_c^{L*}\Omega_p^{L} \right) }{\partial t}\nonumber\\
&+&\Omega_c^{F*}\frac{\partial \Omega_p^F}{\partial y}-\Omega_c^{B*}\frac{\partial \Omega_p^B}{\partial y}
+\Omega_c^{R*}\frac{\partial \Omega_p^R}{\partial x}-\Omega_c^{L*}\frac{\partial \Omega_p^L}{\partial x}\nonumber\\
&=& i \eta \left( \Omega_c^{F*} \rho_{31}^F + 
\Omega_c^{B*} \rho_{31}^B +
\Omega_c^{R*} \rho_{31}^R +
\Omega_c^{L*} \rho_{31}^L  \right) \nonumber\\
&+& \frac{i}{2 k}\left( 
\Omega_c^{F*}\nabla^2\Omega_p^F+
\Omega_c^{B*}\nabla^2\Omega_p^B+
\Omega_c^{R*}\nabla^2\Omega_p^R+
\Omega_c^{L*}\nabla^2\Omega_p^L
\right). 
\end{eqnarray}
when $ \Omega_c^{F\left( B, R, L\right)* }\frac{\partial \Omega_p^{F \left( B, R, L\right)}}{\partial t}\approx\frac{\partial \left[  \Omega_c^{F\left( B, R, L\right)* }\Omega_p^{F \left( B, R, L\right)} \right] }{\partial t}$.
We substitute Eq.(\ref{eq10}-\ref{eq15}) into Eq.~(\ref{eq16}) and get
\begin{eqnarray}
&-&\left[ 1+\frac{V}{c}+\frac{\left( \Delta_c - \Delta_p \right)\left( 2 \Delta_p - i \Gamma\right)  }{\eta c} \right]\frac{\partial \rho_{21}  }{\partial t} +i\left( \Delta_c - \Delta_p \right) \rho_{21} \nonumber\\
&=& \vec{V}_g\cdot \nabla \rho_{21}+\frac{1}{2}\left(\nabla\cdot\vec{V}_g\right)\rho_{21}  \nonumber\\
&-& \frac{ \Gamma + 2 i\Delta_p}{2\eta^2} 
\left[ 
\Omega_c^{F*} \frac{\partial^2 \left( \Omega_c^{F} \rho_{21}\right)}{\partial y^2} 
+\Omega_c^{B*} \frac{\partial^2 \left( \Omega_c^{B} \rho_{21}\right)}{\partial y^2} 
+\Omega_c^{R*} \frac{\partial^2 \left( \Omega_c^{R} \rho_{21}\right) }{\partial x^2}
+\Omega_c^{L*} \frac{\partial^2 \left( \Omega_c^{L} \rho_{21}\right) }{\partial x^2}
\right] \nonumber\\
&-&\frac{i}{4k\eta}\left[ 
\Omega_c^{F*}\nabla^2\left( \Omega_c^{F} \rho_{21}\right) +
\Omega_c^{B*}\nabla^2\left( \Omega_c^{B} \rho_{21}\right) +
\Omega_c^{R*}\nabla^2\left( \Omega_c^{R} \rho_{21}\right) +
\Omega_c^{L*}\nabla^2\left( \Omega_c^{L} \rho_{21}\right) 
\right] \nonumber\\
&+& \frac{i\Gamma - 2 \Delta_p }{4k\eta^2}\left[ 
\Omega_c^{F*} \nabla^2 \frac{\partial \left( \Omega_c^{F} \rho_{21}\right)}{\partial y} 
-\Omega_c^{B*} \nabla^2 \frac{\partial \left( \Omega_c^{B} \rho_{21}\right)}{\partial y} 
+\Omega_c^{R*} \nabla^2 \frac{\partial \left( \Omega_c^{R} \rho_{21}\right)}{\partial x} 
-\Omega_c^{L*} \nabla^2 \frac{\partial \left( \Omega_c^{L} \rho_{21}\right)}{\partial x} 
\right].\nonumber\\\label{eq17}
\end{eqnarray}
when the denominator of Eq.~\eqref{eq10} is a constant.
Here
\begin{eqnarray}
V &=& \frac{\vert \Omega_c^{R}\vert^2+\vert \Omega_c^{L}\vert^2+\vert \Omega_c^{F}\vert^2+\vert \Omega_c^{B}\vert^2}{2\eta}= V_R+V_L+V_F+V_B,\\
\vec{V}_g &=& \frac{1 }{2\eta}\left( \vert \Omega_c^{R}\vert^2 - \vert \Omega_c^{L}\vert^2 ,\vert \Omega_c^{F}\vert^2 - \vert \Omega_c^{B}\vert^2, 0 \right)= \left( V_R - V_L , V_F - V_B, 0\right) .
\end{eqnarray}
Finally, we neglect 
(i) $\frac{V}{c}+\frac{\left( \Delta_c - \Delta_p \right)\left( 2 \Delta_p - i \Gamma\right)  }{\eta c}$ in the first bracket and also
(ii) last bracket where third derivative occurs in Eq.~(\ref{eq17}), and arrive at
\begin{eqnarray}
i\frac{\partial \rho_{21}}{\partial t} &=&
-\frac{V}{2k}\nabla^2\rho_{21}
- \left( 
i \vec{V}_g + \frac{1}{2k} \nabla V
\right) \cdot \nabla \rho_{21}
-\frac{i}{2}\left( \nabla \cdot \vec{V}_g\right) \rho_{21}
\nonumber\\
&-&\left( \frac{2\Delta_p-i\Gamma}{\eta}\right) 
\nabla\cdot\left[
\left( V_R+V_L\right) \frac{\partial \rho_{21}}{\partial x}, 
\left( V_F+V_B\right) \frac{\partial \rho_{21}}{\partial y}
\right]\nonumber\\
&-& \left[ 
\Delta_c-\Delta_p+\frac{1}{4k\eta}\left( 
\Omega_c^{F*}\nabla^2 \Omega_c^{F}+
\Omega_c^{B*}\nabla^2 \Omega_c^{B}+
\Omega_c^{R*}\nabla^2 \Omega_c^{R}+
\Omega_c^{L*}\nabla^2 \Omega_c^{L}
\right) \right] \rho_{21} \nonumber\\
&-&\left( \frac{2\Delta_p -i\Gamma}{2\eta^2}\right) \left( 
\Omega_c^{F*} \frac{\partial^2  \Omega_c^{F}}{\partial y^2}
+\Omega_c^{B*} \frac{\partial^2  \Omega_c^{B}}{\partial y^2}
+\Omega_c^{R*} \frac{\partial^2  \Omega_c^{R}}{\partial x^2}
+\Omega_c^{L*} \frac{\partial^2  \Omega_c^{L}}{\partial x^2}
\right)  \rho_{21} 
.\nonumber\\ \label{eq20}
\end{eqnarray}

\section{Landau Gauge EIT}
When $k \gg 1$ and 
\begin{eqnarray}
\Omega_c^R &=& \frac{\Omega_c}{\sqrt{2}},\\
\Omega_c^L &=& \frac{\Omega_c}{\sqrt{2}},\\
\Omega_c^F &=& \frac{\Omega_c}{\sqrt{2}}\sqrt{1+\frac{x}{L_x}},\\
\Omega_c^B &=& \frac{\Omega_c}{\sqrt{2}}\sqrt{1-\frac{x}{L_x}},\\
\Delta_c &=& \Delta_p - \frac{ \Omega_c^2}{16 \Delta_p L_x^2} x^2, 
\end{eqnarray}
where $\Omega_c$ is some constant Rabi frequency,
we get
\begin{eqnarray}
V_R+V_L &=& \frac{\Omega_c^2}{2\eta},\\
V_F+V_B &=& \frac{\Omega_c^2}{2\eta},\\
\vec{V}_g &=& \frac{\Omega_c^2}{2\eta L_x}\left( 0, x, 0 \right),
\end{eqnarray}
and Eq.~(\ref{eq20}) becomes
\begin{eqnarray}\label{eq29}
i\frac{\partial \rho_{21}}{\partial t} 
&=& 
- \left( \frac{\Omega_c^2 \Delta_p}{\eta^2}\right) \nabla^2 \rho_{21}
- i \frac{\Omega_c^2}{2\eta L_x}\left( 0 ,-x \right) \cdot \nabla \rho_{21}
+ \frac{ \Omega_c^2}{16 \Delta_p L_x^2} x^2 \rho_{21},\nonumber\\
&=&\left[ 
  \left( -i\frac{\Omega_c \sqrt{\Delta_p}}{\eta}\frac{\partial}{\partial x} 
\right)^2 
+ \left( -i\frac{\Omega_c \sqrt{\Delta_p}}{\eta}\frac{\partial}{\partial y} 
+ \frac{ \Omega_c}{4 L\sqrt{\Delta_p}} x\right)^2 \right] \rho_{21} \nonumber\\
&+& i\Gamma \left( \frac{\Omega_c^2}{2\eta^2}\right) \nabla^2 \rho_{21}. 
\end{eqnarray}
Multiplying both side with $\hbar$, Eq.~(\ref{eq29}) becomes an electron-like Schr\"odinger equation:
\begin{eqnarray}\label{eq30}
i\hbar\frac{\partial \rho_{21}}{\partial t} 
&=&\frac{\Omega_c^2 \Delta_p}{\hbar \eta^2}\left[ 
   \left( \frac{\hbar}{i}\frac{\partial}{\partial x} 
\right)^2 
+ \left( \frac{\hbar}{i}\frac{\partial}{\partial y} 
+ \frac{ \hbar \eta}{4 L_x\Delta_p} x\right)^2 \right] \rho_{21} \nonumber\\
&+& i\hbar\Gamma \left( \frac{\Omega_c^2}{2\eta^2}\right) \nabla^2 \rho_{21} \nonumber\\
&=&
\frac{\left( \widehat{P}+e\vec{A}\right) ^2}{2m}\rho_{21}+i\hbar\Gamma \left( \frac{\Omega_c^2}{2\eta^2}\right) \nabla^2 \rho_{21}.
\end{eqnarray}
Here the effective wavefunction, momentum operator, effective vector potential, and effective mass are respectively
\begin{eqnarray}
\psi &=& \rho_{21},\\
\widehat{P} &=& \frac{\hbar}{i}\nabla ,\\
\vec{A} &=& \frac{\hbar \eta^2}{2e\Delta_p \Omega_c^2} \vec{V}_g = \frac{ \hbar \eta}{4 e L_x \Delta_p} \left( 0, x, 0\right),\\
m &=& \frac{\hbar \eta^2}{2\Omega_c^2 \Delta_p}.
\end{eqnarray}
One can then get the effective magnetic field
\begin{equation}
\vec{B}=\nabla\times\vec{A}=\frac{ \hbar \eta}{4 e L_x \Delta_p} \left( 0, 0, 1\right),
\end{equation}
the effective cyclotron frequency
\begin{equation}
\omega_B=\frac{e\vert\vec{B}\vert}{m}=\frac{\Omega_c^2}{2L_x\eta}=\frac{\Omega_c^2}{\xi_x\Gamma},
\end{equation}
and the effective magnetic length
\begin{equation}
\mathit{l}_B=\sqrt{\frac{\hbar}{e\vert\vec{B}\vert}}=2\sqrt{\frac{L_x \Delta_p}{\eta}}=\sqrt{\frac{8 \Delta_p}{\xi_x\Gamma}}L_x.
\end{equation}

\subsection{Non-Hermitian effect}
We calculate the effect of the diffusion term in Eq.~\eqref{eq30} in what follows
\begin{eqnarray}
- i\left( \frac{\Gamma}{2\Delta_p} \right)  \frac{1}{2m}  \langle j \vert   \widehat{P}^2 \vert n\rangle &=& \left( \frac{ - i\Gamma}{4m\Delta_p} \right)  \left(   -\frac{m\hbar\omega_B }{2}   \right)  \langle j \vert   \left( \hat{a}^+ -\hat{a}  \right)^2 \vert n\rangle \nonumber\\
&=& 
\left( \frac{  i\Gamma \hbar\omega_B}{8\Delta_p} \right) \langle j \vert   \left(
\hat{a}^{+2} - \hat{a}^+ \hat{a}  - \hat{a} \hat{a}^+ + \hat{a}^2
\right) \vert n\rangle  \nonumber\\
&=& 
\left( \frac{  i\Gamma \hbar\omega_B}{8\Delta_p} \right) 
\left[ 
\langle j \vert   \left(
\hat{a}^{+2}  + \hat{a}^2
\right) \vert n\rangle 
- \langle n \vert   \left(
 \hat{a}^+ \hat{a}  + \hat{a} \hat{a}^+ 
\right) \vert n\rangle \right] \nonumber \\
&=& 
\left( \frac{  i\Gamma \hbar\omega_B}{8\Delta_p} \right) 
\left[ 
\sqrt{\left( n+1\right) \left( n+2\right) } \langle j \vert   n+2 \rangle 
+
\sqrt{ n \left( n-1\right) } \langle j \vert n-2\rangle 
-  
\left( 2n + 1 \right)\langle j \vert   n \rangle  
\right].
\end{eqnarray}
The first term indicates the  $\vert   n \rangle \rightarrow \vert   n+2 \rangle $ spontaneous transition, the second term depicts  the  $\vert   n \rangle \rightarrow \vert   n-2 \rangle $ spontaneous decay, and the third term shows a dissipation of state $\vert   n \rangle$.  Note that the dissipation rate of state $\vert   n+2 \rangle$ is greater than the   $\vert   n \rangle \rightarrow \vert   n+2 \rangle $  spontaneous transition rate,  and suppresses the latter.  Above derivation shows the origin of $\Delta n=2$  decay in Fig. 3.

\subsection{Time sequence}
\begin{figure}[b]
\includegraphics[width=0.9\textwidth]{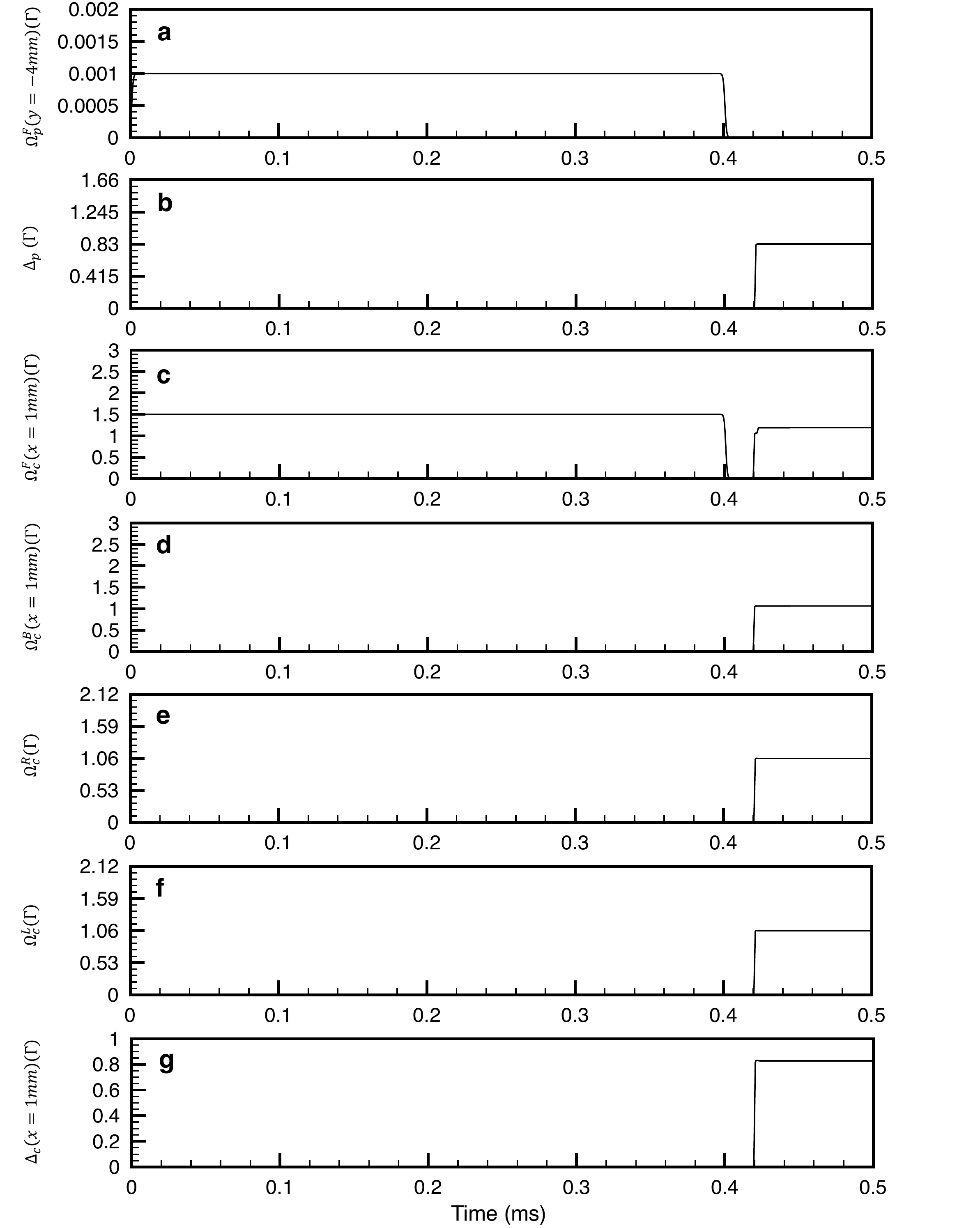}
\caption{\label{figS1}
{\bf Time sequence for Landau-gauge EIT system Figure 3. } 
{\bf a}, Input forward probe Rabi frequency $\Omega_p^F$ at $y=-4$mm where is the  input boundary of the medium.
{\bf b}, Probe detuning  $\Delta_p$  for EIT light  retrieval.  
{\bf c}, Forward control Rabi frequency $\Omega_c^F$ for $x=1$mm.  
{\bf d}, Backward control Rabi frequency $\Omega_c^B$  for $x=1$mm. 
{\bf e}, Rightward control Rabi frequency $\Omega_c^R$. 
{\bf f}, Leftward control Rabi frequency $\Omega_c^L$. 
{\bf g}, Control detuning $\Delta_c$  for $x=1$mm.   
}
\end{figure}

\begin{figure}[b]
\includegraphics[width=0.9\textwidth]{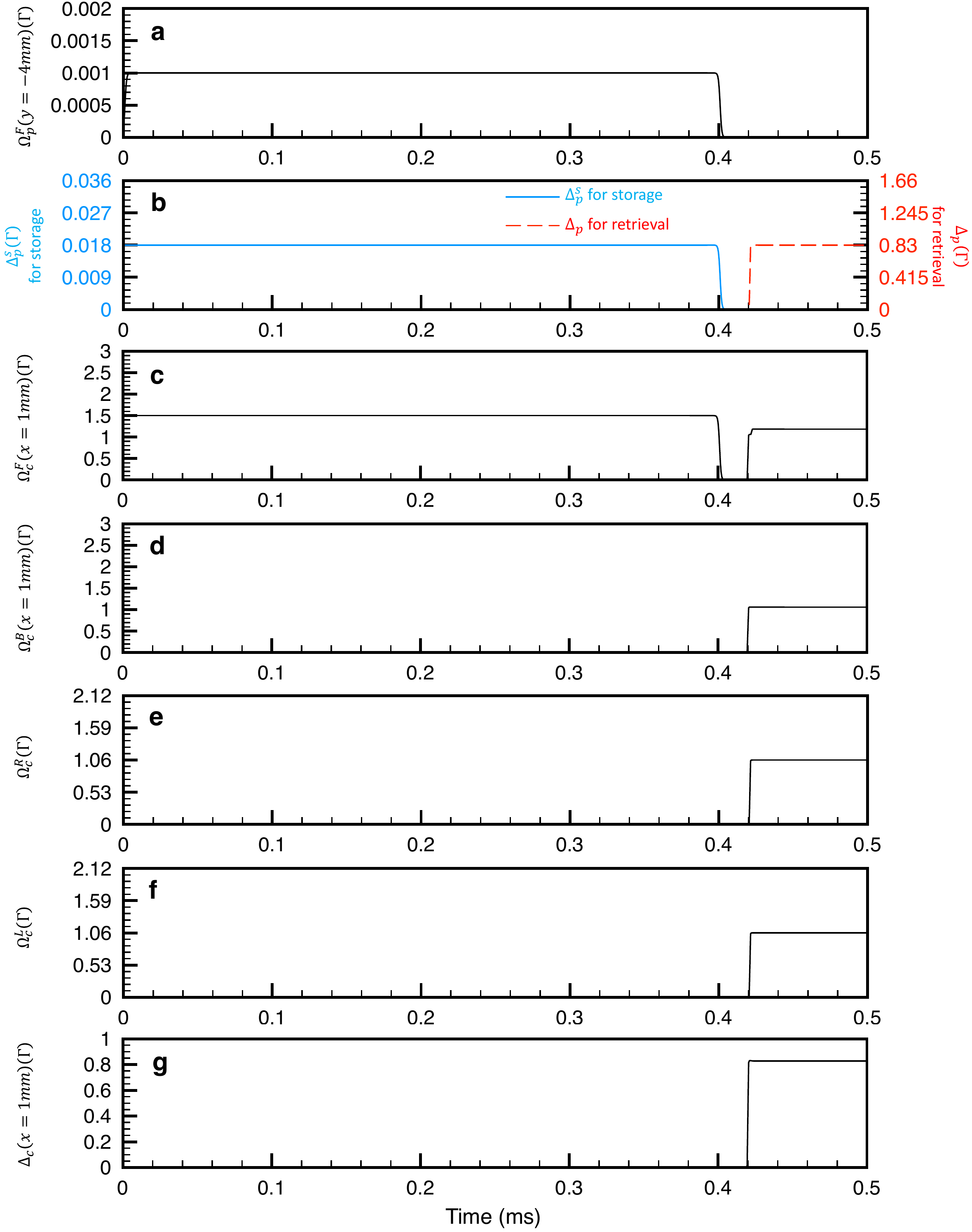}
\caption{\label{figS2}
{\bf Time sequence for Landau-gauge EIT system Figure 4. } 
{\bf a}, Input forward probe Rabi frequency $\Omega_p^F$ at $y=-4$mm where is the  input boundary of the medium.
{\bf b}, Probe detuning $\Delta_p^s$ and $\Delta_p$ are for EIT light storage (blue-solid line) and retrieval (red-dashed line), respectively.  $\Delta_p^s=-0.018\Gamma$ is for Fig.4(a,d), $\Delta_p^s=0$ for Fig.4(b,e), and $\Delta_p^s=0.018\Gamma$ is for Fig.4(c,f).
{\bf c}, Forward control Rabi frequency $\Omega_c^F$ for $x=1$mm.  
{\bf d}, Backward control Rabi frequency $\Omega_c^B$   for $x=1$mm. 
{\bf e}, Rightward control Rabi frequency $\Omega_c^R$. 
{\bf f}, Leftward control Rabi frequency $\Omega_c^L$. 
{\bf g}, Control detuning $\Delta_c$  for $x=1$mm. 
}
\end{figure}

The time sequences used in our numerical simulation of Eqs.~(\ref{eq1}-\ref{eq9}) for Fig. 3 and Fig.4 are demonstrated in Fig.~\ref{figS1} and Fig.~\ref{figS2}, respectively. The simulation box is in a two-dimensional domain of -4.5mm$ \leq x\leq$ 4.5mm and -4mm$ \leq y\leq$ 4mm.  The boundary condition at $y=-4$mm is 
\begin{equation}
\Omega_p^F\left( x,  y = -4mm, t \right) = \Omega_0\left( t\right)  H_n\left(  \frac{x + k\mathit{l}_B^2}{\mathit{l}_B}\right) e^{ -\left(  \frac{x + k\mathit{l}_B^2}{\sqrt{2}\mathit{l}_B}\right)^2},\nonumber
\end{equation}
where $\Omega_0\left( t\right)$ is demonstrated in Fig.~\ref{figS1}\textbf{a}  and Fig.~\ref{figS2}\textbf{a}.
The $\Delta_p^s = 0.018\Gamma$ in Fig.~\ref{figS2}\textbf{b} is for $k_s = -\frac{2\pi}{L_y} $ in Fig.4\textbf{c}\&\textbf{f}.  
$\Delta_p^s = -0.018\Gamma$ and 0 are used for $k_s = \frac{2\pi}{L_y} $ and 0 in Fig.4, respectively.
The Rabi frequency of control fields   in Fig.~\ref{figS1}\textbf{c}-\textbf{f}   and Fig.~\ref{figS2}\textbf{c}-\textbf{f} are given by
\begin{eqnarray}
\Omega_c^F\left( x, t  \right) &=& 
\frac{\Omega_c}{2}\left[ 1-\tanh\left( \frac{t-t_s}{\tau_s/4}\right)\right] \nonumber\\
&+&\frac{\Omega_c}{2\sqrt{2}}\left[ 1+\tanh\left( \frac{t-t_r}{\tau/4}\right) \right] \left\lbrace  1 + \left( -1+ \frac{\Omega_c}{\sqrt{2}}\sqrt{1+\frac{x}{L_x}} \right)\frac{1}{2}\left[ 1+\tanh\left( \frac{t-t_{LG}}{\tau/4}\right) \right] \right\rbrace , \nonumber\\
 \Omega_c^B\left( x, t  \right) &=& \frac{\Omega_c}{2\sqrt{2}}\left[ 1+\tanh\left( \frac{t-t_r}{\tau/4}\right) \right] \left\lbrace  1 + \left( -1+ \frac{\Omega_c}{\sqrt{2}}\sqrt{1-\frac{x}{L_x}} \right)\frac{1}{2}\left[ 1+\tanh\left( \frac{t-t_{LG}}{\tau/4}\right) \right] \right\rbrace, \nonumber\\
 \Omega_c^R\left( x, t  \right) &=& \frac{\Omega_c}{2\sqrt{2}}\left[ 1+\tanh\left( \frac{t-t_r}{\tau/4}\right) \right] ,\nonumber\\
 \Omega_c^L\left( x, t  \right) &=& \frac{\Omega_c}{2\sqrt{2}}\left[ 1+\tanh\left( \frac{t-t_r}{\tau/4}\right) \right],\nonumber
\end{eqnarray}
whose peak value is $\frac{\Omega_c}{\sqrt{2}}=\frac{1.5\Gamma}{\sqrt{2}}=1.06\Gamma$.
The detuning of control fields is 
\begin{equation}
\Delta_c\left( x, t  \right) = \frac{\Delta_p}{2}\left[ 1+\tanh\left( \frac{t-t_r}{\tau/4}\right) \right] -
\frac{ \Omega_c^2}{16 \Delta_p L_x^2} x^2 \frac{1}{2}\left[ 1+\tanh\left( \frac{t-t_{LG}}{\tau/4}\right) \right]. \nonumber\\
\end{equation}
Here $t_s=0.4$ms, $t_r=0.42$ms, $t_{LG}=0.4225$ms, $\tau_s= 4\mu$s, and $\tau= 1\mu$s.
\begin{figure}[b]
\includegraphics[width=0.9\textwidth]{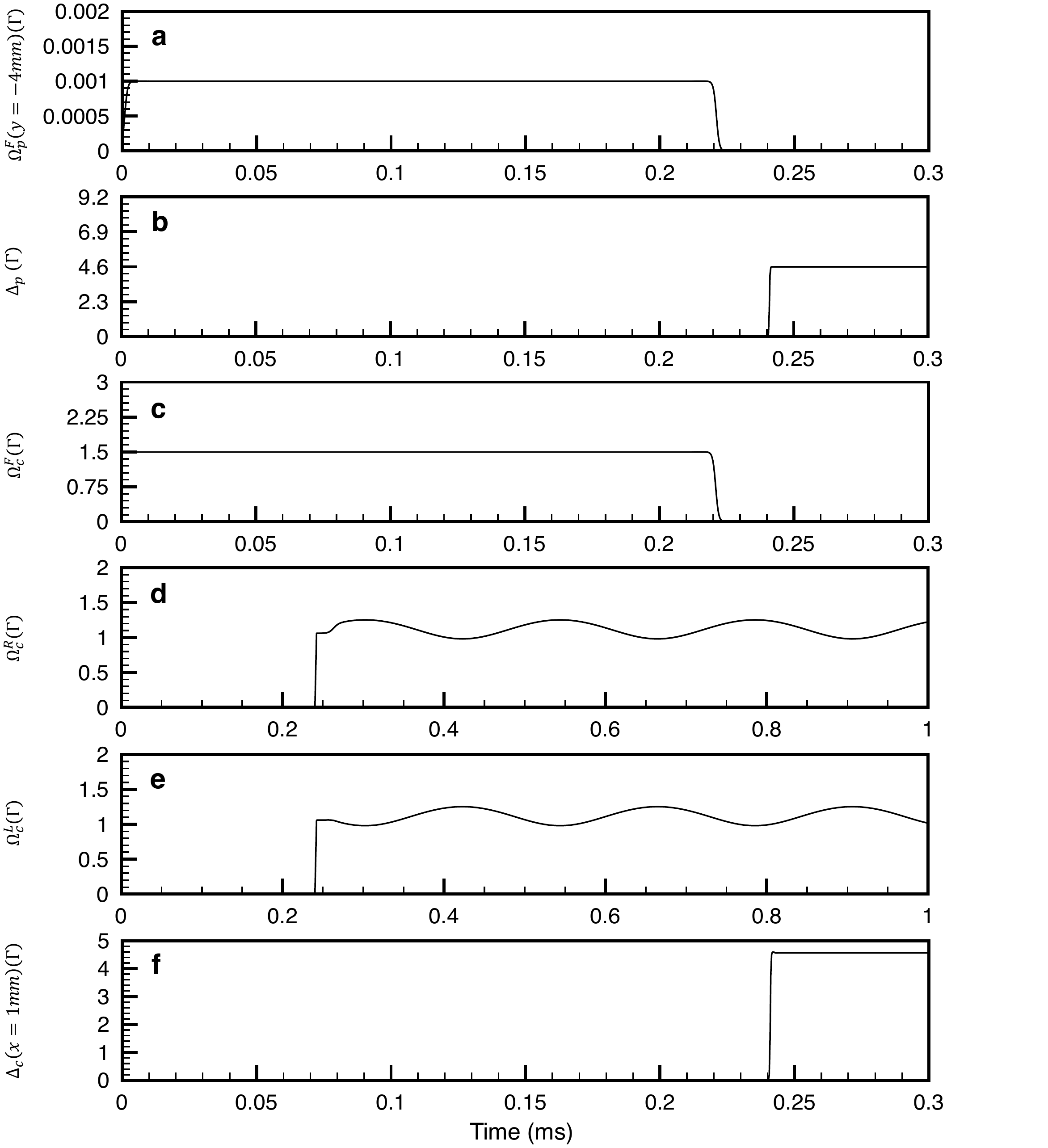}
\caption{\label{figS3}
{\bf Time sequence for the driven QHO system Figure 5 } 
{\bf a}, Input forward probe Rabi frequency $\Omega_p^F$ at $y=-4$mm where is the  input boundary of the medium.
{\bf b}, Probe detuning $\Delta_p$.
{\bf c}, Forward control Rabi frequency $\Omega_c^F$.  
{\bf d}, Rightward control Rabi frequency $\Omega_c^R$.
{\bf f}, Leftward control Rabi frequency $\Omega_c^L$.
{\bf g}, Control detuning $\Delta_c$  for $x=1$mm.  
}
\end{figure}

\section{Driving quantum harmonic oscillator}
In order to simulate Rabi oscillation of a two-level quantum harmonic oscillator in a 1D EIT system, we introduce $\hat{P}\cdot\vec{A}$ Hamiltonian and perturbed quadratic potential by using
\begin{eqnarray}
\Omega_c^R &=& \frac{\Omega_c}{\sqrt{2}}\sqrt{1+\alpha\sin\left( \omega_d t\right)},\\
\Omega_c^L &=& \frac{\Omega_c}{\sqrt{2}}\sqrt{1-\alpha\sin\left( \omega_d t\right)},\\
\Delta_c &=& \Delta_p  - \frac{1}{2}\omega_E \left(\frac{x}{\mathit{l}_E}\right)^2-\frac{2}{3}\beta\omega_E \left(\frac{x}{\mathit{l}_E}\right)^4-\alpha^2\frac{m\Omega_c^4}{8\hbar\eta^2}\sin^2\left( \omega_d t\right), 
\end{eqnarray}
where  the characteristic QHO length $\mathit{l}_E=\sqrt{\frac{\hbar}{m\omega_E}}$.
We get
\begin{eqnarray}
V_R+V_L &=& \frac{\Omega_c^2}{2\eta},\\
\vec{V}_g &=& \left( \alpha\frac{\Omega_c^2}{2\eta } \sin\left( \omega_d t\right), 0, 0 \right),
\end{eqnarray}
and Eq.~(\ref{eq20}) becomes
\begin{eqnarray}\label{eq43}
i\frac{\partial \rho_{21}}{\partial t} 
&=& 
- \hbar\left( \frac{\Delta_p\Omega_c^2 }{\hbar\eta^2}\right) \partial_x^2 \rho_{21}
- i\alpha \frac{\Omega_c^2}{2\eta }\sin\left( \omega_d t\right) \partial_x \rho_{21}
\nonumber\\
&+& \left[    \frac{1}{2\hbar}m\omega_E^2 x^2 +\beta\frac{2 m^2 \omega_E^3}{3\hbar^2} x^4  +\alpha^2\frac{m\Omega_c^4}{8\hbar\eta^2}\sin^2\left( \omega_d t\right)\right] \rho_{21} + i\Gamma \left( \frac{\Omega_c^2}{2\eta^2}\right) \partial_x^2 \rho_{21} \nonumber\\
&=& \frac{1}{2m} \left[ \hbar\left( -i \partial_x \right) ^2 + 2m\alpha \frac{\Omega_c^2}{2\eta }\sin\left( \omega_d t\right) \left( - i\partial_x\right) 
+m^2\alpha^2\frac{\Omega_c^4}{4\hbar\eta^2}\sin^2\left( \omega_d t\right)
\right] \rho_{21} \nonumber\\
&+& \left(    \frac{1}{2\hbar}m\omega_E^2 x^2 +\beta\frac{2 m^2 \omega_E^3}{3\hbar^2} x^4  \right) \rho_{21} + i\frac{\hbar\Gamma}{2\Delta_p} \left( \frac{\Delta_p\Omega_c^2}{\hbar\eta^2}\right) \partial_x^2 \rho_{21}. 
\end{eqnarray}
Multiplying both side with $\hbar$, Eq.~(\ref{eq43}) becomes an electron-like Schr\"odinger equation:
\begin{eqnarray}\label{eq44}
i\hbar\frac{\partial \rho_{21}}{\partial t} 
&=& \frac{1}{2m} \left[ \left( -i \hbar\partial_x \right) ^2 + 2m\alpha \frac{\Omega_c^2}{2\eta }\sin\left( \omega_d t\right) \left( - i\hbar\partial_x\right) 
+\alpha^2\frac{m^2\Omega_c^4}{4\eta^2}\sin^2\left( \omega_d t\right)
\right] \rho_{21} \nonumber\\
&+& \left(    \frac{1}{2}m\omega_E^2 x^2 +\beta\frac{2 m^2 \omega_E^3}{3\hbar} x^4  \right) \rho_{21} - i\frac{\Gamma}{2\Delta_p} \left( \frac{1}{2m}\right) \left( -i \hbar\partial_x\right) ^2 \rho_{21}. \nonumber\\
&=& \frac{1}{2m} \left\lbrace\left( -i \hbar\partial_x \right) + e \left[  \alpha \frac{m\Omega_c^2}{2e\eta }\sin\left( \omega_d t\right)\right]  \right\rbrace ^2 \rho_{21} \nonumber\\
&+& \left(    \frac{1}{2}m\omega_E^2 x^2 +\beta\frac{2 m^2 \omega_E^3}{3\hbar} x^4  \right) \rho_{21} - i\frac{\Gamma}{2\Delta_p} \left( \frac{1}{2m}\right) \left( -i \hbar\partial_x\right) ^2 \rho_{21}. \nonumber\\
&=&
\frac{\left( \widehat{P}+e\vec{A}\right) ^2}{2m}\rho_{21}+U \rho_{21} - i\left( \frac{\Gamma}{2\Delta_p} \right)  \frac{\widehat{P}^2}{2m}  \rho_{21}.
\end{eqnarray}
Here the effective wavefunction, momentum operator, effective vector potential, and effective mass are respectively
\begin{eqnarray}
\psi &=& \rho_{21},\\
\widehat{P} &=& \frac{\hbar}{i}\partial_x ,\\
\vec{A} &=& \frac{\hbar \eta^2}{2e\Delta_p \Omega_c^2} \vec{V}_g =  \left( \alpha \frac{\hbar\eta}{4e\Delta_p }\sin\left( \omega_d t\right), 0, 0\right),\\
U &=&  \frac{1}{2}m\omega_E^2 x^2 +\beta\frac{2 m^2\omega_E^3}{3\hbar} x^4,\\
m &=& \frac{\hbar \eta^2}{2\Omega_c^2 \Delta_p}.
\end{eqnarray}
One can then get the effective electric field and magnetic field
\begin{eqnarray}
\vec{E} &=& -\nabla U-\frac{\partial \vec{A}}{\partial t} =  \left( - m\omega_E^2 x -\beta\frac{8 m^2\omega_E^3}{3\hbar} x^3-\alpha\frac{ \hbar \eta \omega_d}{4 e  \Delta_p}\cos\left( \omega_d t\right), 0, 0\right), \\
\vec{B} &=& \nabla\times\vec{A}=0.
\end{eqnarray}

\subsection{Driving Rabi frequency}
We  calculate the effective Rabi frequency $\Omega_A=\frac{e}{m\hbar}\vec{A} \cdot  \hat{P}$ and neglect the small $\vert \vec{A}\vert^2$ term as typically adopted in quantum optics
\begin{eqnarray}
\Omega_A &=& \frac{e}{m\hbar}\vec{A} \cdot\langle n+1\vert \hat{P} \vert n\rangle =   \frac{\alpha\eta}{4 m \Delta_p } \sqrt{\frac{\hbar m \omega_E}{2}} \langle n+1\vert \left( \hat{a}^+ -\hat{a}  \right)  \vert n\rangle 
= \frac{\alpha\eta}{4 \Delta_p } \sqrt{\frac{\hbar \omega_E \left( n+1\right)}{2m}} \nonumber\\
&=& \frac{\alpha\Omega_c}{4} \sqrt{\frac{ \omega_E \left( n+1\right)}{\Delta_p}},
\end{eqnarray}
where $\hat{a}^+$ and $\hat{a}$ are QHO raising and lowering operators, respectively. 
In the main text we use 
\begin{equation}
\Delta_c = \Delta_p  - \frac{1}{2}\omega_E \left(\frac{x}{\mathit{l}_E}\right)^2-\frac{2}{3}\beta\omega_E \left(\frac{x}{\mathit{l}_E}\right)^4,
\end{equation}
which also neglects the  $\vert \vec{A}\vert^2$ term and leads to the same Rabi frequency.  

\subsection{The first-order perturbed eigen-energy}
We invoke raising and lowering operators to calculate the first-order perturbed energy by the quartic term.
\begin{eqnarray}
\delta E_{n}^1 &=& \beta\frac{2 m^2\omega_E^3}{3\hbar} \langle n \vert x^4 \vert n\rangle \nonumber\\
&=& \beta\frac{2 m^2\omega_E^3}{3\hbar} \left(  \frac{\hbar}{2m\omega_E}   \right)^2 \langle n \vert  \left( \hat{a}^+ +\hat{a}  \right)^4 \vert n\rangle \nonumber\\
&=& \frac{\beta}{6}\hbar \omega_E  \langle n \vert  \left(   
\hat{a}^{+2}\hat{a}^2 + \hat{a}^+ \hat{a} \hat{a}^+ \hat{a} +  \hat{a}^+ \hat{a}^2 \hat{a}^+ + \hat{a} \hat{a}^{+2} \hat{a} + \hat{a} \hat{a}^+  \hat{a} \hat{a}^+ + \hat{a}^2 \hat{a}^{+2} 
\right) \vert n\rangle \nonumber\\
&=& \beta \left( n^2 + n + \frac{1}{2} \right) \hbar\omega_E .
\end{eqnarray}
The first-order perturbed eigen energy under the  quartic term reads
\begin{equation}
E_n = \left( n + \frac{1}{2} \right) \hbar\omega_E + \beta \left( n^2 + n + \frac{1}{2} \right) \hbar\omega_E .
\end{equation}

\subsection{Time sequence}
The time sequences used in our numerical simulation of Eqs.~(\ref{eq1}-\ref{eq9}) for Fig. 5 are demonstrated in Fig.~\ref{figS3}. The simulation box is in a two-dimensional domain of -4mm$ \leq x\leq$ 4mm and -4mm$ \leq y\leq$ 4mm.  The boundary condition at $y=-4$mm is 
\begin{equation}
\Omega_p^F\left( x,  y = -4mm, t \right) = \Omega_0\left( t\right) H_n\left(  \frac{x}{\mathit{l}_E}\right) e^{-\left(  \frac{x}{\sqrt{2}\mathit{l}_E}\right)^2},\nonumber
\end{equation}
where $\Omega_0\left( t\right)$ is demonstrated in Fig.~\ref{figS3}\textbf{a}.
The Rabi frequency of control fields  in Fig.~\ref{figS3}\textbf{d}\&\textbf{e}  are given by
\begin{eqnarray}
\Omega_c^F\left( x, t  \right) &=& 
\frac{\Omega_c}{2}\left[ 1-\tanh\left( \frac{t-t_s}{\tau_s/4}\right)\right], \nonumber\\
\Omega_c^R\left( x, t  \right) &=& \frac{\Omega_c}{2\sqrt{2}}\left[ 1+\tanh\left( \frac{t-t_r}{\tau/4}\right) \right] \left\lbrace  1 + \left( -1+ \frac{\Omega_c}{\sqrt{2}}\sqrt{1+\alpha\sin\left( \omega_d t\right)} \right)\frac{1}{2}\left[ 1+\tanh\left( \frac{t-t_{D}}{\tau_D/4}\right) \right] \right\rbrace , \nonumber\\
 \Omega_c^L\left( x, t  \right) &=& \frac{\Omega_c}{2\sqrt{2}}\left[ 1+\tanh\left( \frac{t-t_r}{\tau/4}\right) \right] \left\lbrace  1 + \left( -1+ \frac{\Omega_c}{\sqrt{2}}\sqrt{1-\alpha\sin\left( \omega_d t\right)} \right)\frac{1}{2}\left[ 1+\tanh\left( \frac{t-t_{D}}{\tau_D/4}\right) \right] \right\rbrace, \nonumber
\end{eqnarray}
whose peak value is $\frac{\Omega_c}{\sqrt{2}}=\frac{1.5\Gamma}{\sqrt{2}}=1.06\Gamma$.
The detuning of control fields is 
\begin{equation}
\Delta_c\left( x, t  \right) = \frac{\Delta_p}{2}\left[ 1+\tanh\left( \frac{t-t_r}{\tau/4}\right) \right] -
\left[    \frac{1}{2}\omega_E \left(\frac{x}{\mathit{l}_E}\right)^2+\frac{2}{3}\beta\omega_E \left(\frac{x}{\mathit{l}_E}\right)^4     \right]  \frac{1}{2}\left[ 1+\tanh\left( \frac{t-t_{H}}{\tau/4}\right) \right]. \nonumber\\
\end{equation}
Here $t_s=0.22$ms, $t_r=0.241$ms, $t_{H}=0.2425$ms, $t_{D}=0.2625$ms, $\tau_s=4 \mu$s, $\tau=1 \mu$s, and $\tau_D= 25\mu$s.

\end{document}